\documentclass[12pt]{article}
\usepackage{amssym}

\font\elevenof=msbm10 at 11pt
\font\seventeenof=msbm10 at 17pt

\def\Z{\mbox{$\Bbb Z$}}
\def\R{\mbox{$\Bbb R$}}
\def\N{\mbox{$\Bbb N$}}
\def\AG{${\cal A}(G(N))$}
\def\ap{a^{\dagger}}

\def\mod{\mathop{\rm mod}\nolimits}
\def\diag{\mathop{\rm diag}\nolimits}
\def\case#1#2{{\textstyle{#1\over #2}}}

\title{
\hfill{\normalsize ULB/229/CQ/02/9}\\
\vspace{1cm} 
FRACTIONAL SUPERSYMMETRIC QUANTUM MECHANICS, TOPOLOGICAL INVARIANTS AND
GENERALIZED DEFORMED OSCILLATOR ALGEBRAS}
\author{C. QUESNE\thanks{Directeur de recherches FNRS}\\
{\small \sl Physique Nucl\'eaire Th\'eorique et Physique Math\'ematique,}\\ {\small \sl
Universit\'e Libre de Bruxelles, Campus de la Plaine CP229,} \\ {\small \sl  Boulevard~du
Triomphe, B-1050 Brussels, Belgium} \\
{\small \sl E-mail: cquesne@ulb.ac.be}}
\date{ }
\begin{document}
\baselineskip=22pt plus 1pt minus 1pt
\maketitle

\begin{abstract} 
Fractional supersymmetric quantum mechanics of order $\lambda$ is realized in terms of
the generators of a generalized deformed oscillator algebra and a $\mbox{\elevenof
Z}_{\lambda}$-grading structure is imposed on the Fock space of the latter. This
realization is shown to be fully reducible with the irreducible components providing
$\lambda$ sets of minimally bosonized operators corresponding to both unbroken and
broken cases. It also furnishes some examples of $\mbox{\elevenof
Z}_{\lambda}$-graded uniform topological symmetry of type (1, 1, \ldots, 1) with
topological invariants generalizing the Witten index. 
\end{abstract}

\noindent
Running head: Fractional Supersymmetric Quantum Mechanics

\noindent
Keywords: Quantum mechanics; fractional supersymmetry; topological invariants;
generalized deformed oscillator

\noindent
PACS Nos.: 03.65.Fd, 11.30.Pb
%
%
\section{Introduction}

Supersymmetric quantum mechanics (SSQM), which was originally introduced as a testing
bench for some new ideas in quantum field theory~\cite{witten}, has found a lot of
applications in various fields (for reviews see e.g.~\cite{cooper}). In such a theory, there
is a $\Z_2$-grading of the Hilbert space and the Hamiltonian $H$ is written as the square
of (at least) one conserved supercharge: $Q^2 = H$, with $[H, Q] = 0$.\par
%
%
Since its introduction, it has been extended in various ways, thus giving rise for instance
to parasupersymmetric (PSSQM)~\cite{rubakov, beckers90},
orthosupersymmetric\linebreak (OSSQM)~\cite{khare},
pseudosupersymmetric~\cite{beckers95}, and fractional supersymmetric quantum
mechanics (FSSQM)~\cite{ahn, baulieu, kerner, filippov, durand, moha, fleury, azca}. In this
letter, our main interest will be the latter, where one replaces the $\Z_2$-grading
characterizing SSQM by a $\Z_{\lambda}$-grading in such a way that the Hamiltonian
becomes the $\lambda$th power of a conserved fractional supercharge:
\begin{equation}
  Q^{\lambda} = H,  \label{eq:def-1}
\end{equation}
with
\begin{equation}
  [H, Q] = 0  \label{eq:def-2}
\end{equation}
and $\lambda \in \{3, 4, 5, \ldots\}$.\par
%
%
It is usual to realize FSSQM of order $\lambda$ in terms of bosonic creation and
annihilation operators together with some operators generalizing the fermionic ones. The
latter , which are distinct from the parafermionic operators of order
$\lambda$~\cite{green}, are related instead to the $q$-deformed harmonic
oscillator~\cite{arik} with $q$ a primitive $\lambda$th root of unity, e.g., $q = \exp(2\pi
{\rm i}/\lambda)$ such that $q^{\lambda} =1$.\par
%
%
Here, in line with our previous studies of some other variants of SSQM~\cite{cq98, cq02},
we shall adopt another viewpoint and realize FSSQM of order $\lambda$ in terms of
generalized deformed oscillator algebra (GDOA) generators (\cite{katriel} and references
quoted therein) by imposing a $\Z_{\lambda}$-grading structure on the corresponding
Fock space. As a result, FSSQM of order $\lambda$ will prove fully reducible and we shall
get a {\em minimal\/} bosonization of this theory in terms of a {\em single\/} bosonic
degree of freedom.\par
%
%
Another purpose of this letter is connected with the concept of topological symmetries,
which has recently been introduced~\cite{mosta} in an attempt to construct
generalizations of supersymmetry sharing its topological properties in the sense that they
involve integer-valued topological invariants similar to the Witten index~\cite{witten}.
Since a special case of $\Z_{\lambda}$-graded topological symmetries has been shown
to be related to FSSQM of order $\lambda$~\cite{mosta}, our realization of the latter in
terms of GDOA generators will provide us with simple examples of the former and of its
topological invariants.\par
%
%
\section{Generalized Deformed Oscillator Algebras}

Let us start with a brief review of GDOAs~\cite{katriel}.\par
%
%
A GDOA may be defined as a nonlinear associative algebra \AG\ generated by the
operators $N = N^{\dagger}$, $\ap$, and $a = (\ap)^{\dagger}$, satisfying the
commutation relations
\begin{equation}
  [N, \ap] = \ap, \qquad [N, a] = - a, \qquad [a, \ap] = G(N),  \label{eq:GDOA}
\end{equation}
where $G(N) = [G(N)]^{\dagger}$ is some Hermitian function of $N$.\par
%
%
We restrict ourselves here to GDOAs possessing a bosonic Fock space representation. In
the latter, we may write $\ap a = F(N)$, $a \ap = F(N+1)$, where the structure function
$F(N) = [F(N)]^{\dagger}$ is such that
\begin{equation}
  G(N) = F(N+1) - F(N) \label{eq:G-F}
\end{equation}
and is assumed to satisfy the conditions
\begin{equation}
  F(0) = 0, \qquad F(n) > 0 \qquad \mbox{\rm if\ } n = 1, 2, 3, \ldots.  \label{eq:F} 
  \label{eq:F-cond}  
\end{equation}
The carrier space $\cal F$ of such a representation can be constructed from a vacuum
state $|0\rangle$ (such that $a |0\rangle = N |0\rangle = 0$) by successive applications
of the creation operator $\ap$. Its basis states
\begin{equation}
  |n\rangle = \left(\prod_{i=1}^n F(i)\right)^{-1/2} (\ap)^n |0\rangle, \qquad n=0, 1, 2,
  \ldots,  \label{eq:basis}
\end{equation}
where we set $\prod_{i=1}^0 \equiv 1$, satisfy the relations $N |n\rangle = n |n\rangle$,
$\ap |n\rangle = \sqrt{F(n+1)} |n+1\rangle$, and $a |n\rangle = \sqrt{F(n)}
|n-1\rangle$.\par
%
%
{}For $G(N) = I$, we obtain $F(N) = N$ and the algebra \AG\ reduces to the standard
(bosonic) oscillator algebra ${\cal A}(I)$, for which the creation and annihilation operators
may be written as $\ap = (x - {\rm i}P)/\sqrt{2}$, $a = (x + {\rm i}P)/\sqrt{2}$, where
$P$ denotes the momentum operator ($P = - {\rm i} d/dx$).\par
%
%
A $\Z_{\lambda}$-grading structure can be imposed on $\cal F$ by introducing a grading
operator
\begin{equation}
  T = e^{2\pi{\rm i}N/\lambda}, \qquad \lambda \in \{2, 3, 4, \ldots\},  \label{eq:T} 
\end{equation}
which is such that
\begin{equation}
  T^{\dagger} = T^{-1}, \qquad T^{\lambda} = I.  \label{eq:T-prop1}
\end{equation}
It has $\lambda$ distinct eigenvalues $q^{\mu}$, $\mu = 0$, 1,~\ldots, $\lambda-1$,
with corresponding eigenspaces ${\cal F}_{\mu}$ spanned by $|n\rangle = |k\lambda +
\mu \rangle$, $k=0$, 1, 2, \ldots, and such that ${\cal F} = \sum_{\mu=0}^{\lambda-1}
\oplus {\cal F}_{\mu}$. Here $q \equiv \exp(2\pi {\rm i}/\lambda)$.
From~(\ref{eq:GDOA}), it results that
$T$ satisfies the relations
\begin{equation}
  [N, T] = 0, \qquad \ap T = q^{-1} T \ap, \qquad a T = q T a,  \label{eq:T-prop2} 
\end{equation}
expressing the fact that $N$ preserves the grade, while $\ap$ (resp.\ $a$) increases
(resp.\ decreases) it by one unit.\par
%
%
The operators
\begin{equation}
  P_{\mu} = \frac{1}{\lambda} \sum_{\nu=0}^{\lambda-1} q^{-\mu\nu} T^{\nu},
  \qquad \mu=0, 1, \ldots, \lambda-1,  \label{eq:P}
\end{equation}
project on the various subspaces ${\cal F}_{\mu}$, $\mu=0$, 1,~\ldots, $\lambda-1$,
and therefore satisfy the relations
\begin{equation}
  P_{\mu}^{\dagger} = P_{\mu}, \qquad P_{\mu} P_{\nu} = \delta_{\mu,\nu} P_{\mu},
  \qquad \sum_{\mu=0}^{\lambda-1} P_{\mu} = I  \label{eq:P-prop1}
\end{equation}
in $\cal F$. As a consequence of~(\ref{eq:T-prop2}), they also fulfil the relations
\begin{equation}
  [N, P_{\mu}] = 0, \qquad \ap P_{\mu} = P_{\mu+1} \ap, \qquad a P_{\mu} =
  P_{\mu-1} a,  \label{eq:P-prop2}
\end{equation}
where we use the convention $P_{\mu'} = P_{\mu}$ if $\mu' - \mu = 0 \mod \lambda$.
\par
%
%
A special case of GDOA with a built-in $\Z_{\lambda}$-grading structure is provided by
the GDOA associated with a $C_{\lambda}$-extended oscillator algebra
${\cal A}^{(\lambda)}_{\alpha_0 \alpha_1 \ldots \alpha_{\lambda-2}}$, where the
cyclic group $C_{\lambda} = \Z_{\lambda}$ is generated by $T$, i.e., $C_{\lambda} =
\{T, T^2, \ldots, T^{\lambda-1}, T^{\lambda} = I\}$ \cite{cq98}. It corresponds to the
choice $G(N) = I + \sum_{\mu=0}^{\lambda-1} \alpha_{\mu} P_{\mu}$, where
$\alpha_{\mu}$, $\mu=0$, 1,~\ldots, $\lambda-1$, are some real parameters
constrained by $\sum_{\mu=0}^{\lambda-1} \alpha_{\mu} = 0$, and it reduces to the
Calogero-Vasiliev algebra~\cite{vasiliev} in the $\lambda = 2$ limit.\par
%
%
\section{Fractional Supersymmetric Quantum Mechanics}

Let us now look for a $\lambda \times \lambda$-matrix realization of the fractional
supercharge and supersymmetric Hamiltonian of the type 
\begin{equation}
  Q = \sum_{i=1}^{\lambda-1} A_i e_{i+1,i} + A_{\lambda} e_{1,\lambda}, \qquad H = 
  \sum_{i=1}^{\lambda} h_i e_{i,i},  \label{eq:Q-H} 
\end{equation}
where
\begin{eqnarray}
  A_i & = & f_i(N+i) a, \qquad i=1, 2, \ldots, \lambda-1, \qquad A_{\lambda} = 
        f_{\lambda}(N) (\ap)^{\lambda-1}, \label{eq:A} \\
  h_i & = & h_i(N), \qquad i=1, 2, \ldots, \lambda, \label{eq:h}
\end{eqnarray}
are defined in terms of the generators $N$, $\ap$, $a$ of a GDOA \AG. Here $e_{i,j}$
denotes the $\lambda$-dimensional matrix with entry 1 at the intersection of row $i$
and column $j$ and zeros everywhere else, while $f_i(N)$ and $h_i(N)$, $i=1$, 2,
\ldots, $\lambda$, are some complex and real functions of $N$, respectively. The
functions $f_i(N)$ are furthermore restricted by the condition that
\begin{equation}
  \varphi(N) \equiv \prod_{i=1}^{\lambda} f_i(N)  \label{eq:phi}
\end{equation}
be such that
\begin{equation}
  \varphi(n) \in \R^+ \qquad \mbox{\rm if\ } n = \lambda-1, \lambda, \lambda+1, \ldots. 
  \label{eq:phi(n)}
\end{equation}
\par
%
%
On inserting (\ref{eq:Q-H}) into (\ref{eq:def-1}), we obtain $\lambda$ conditions
\begin{eqnarray}
  A_{\lambda} A_{\lambda-1} \ldots A_1 & = & h_1, \nonumber \\
  A_{i-1} A_{i-2} \ldots A_1 A_{\lambda} A_{\lambda-1} \ldots A_i & = & h_i, \qquad
        i=2, 3, \ldots, \lambda,
\end{eqnarray}
which, on taking (\ref{eq:A}) and (\ref{eq:h}) into account, reduce to
\begin{equation}
  h_i(N) = h_1(N+i-1) = \varphi(N+i-1) \prod_{j=1}^{\lambda-1} F(N+i-j), \qquad i=2, 3,
  \ldots, \lambda.  \label{eq:h-bis}
\end{equation}
It is then straightforward to check that with this choice, $H$ and $Q$ also satisfy
Eq.~(\ref{eq:def-2}). Hence $H$ is completely determined by the function $\varphi(N)$,
defined in (\ref{eq:phi}), and by the GDOA structure function $F(N)$.\par
%
%
In FSSQM, it is well known that there exists another conserved fractional supercharge, the
fractional covariant derivative $D$ (see e.g.~\cite{durand, azca}). It satisfies relations
similar to (\ref{eq:def-1}) and (\ref{eq:def-2}),
\begin{equation}
  D^{\lambda} = H, \qquad [H, D] = 0,  \label{eq:D-1}
\end{equation}
as well as a $q$-commutation relation with $Q$,
\begin{equation}
  [D, Q]_q \equiv DQ - qQD = 0.  \label{eq:D-2}
\end{equation}
\par
%
%
A $\lambda \times \lambda$-matrix realization of $D$ can be obtained in the form
\begin{equation}
  D = \sum_{i=1}^{\lambda-1} B_i e_{i+1,i} + B_{\lambda} e_{1,\lambda},  \label{eq:D}
\end{equation}
where
\begin{equation}
  B_i = g_i(N+i) a, \qquad i=1, 2, \ldots, \lambda-1, \qquad B_{\lambda} = 
        g_{\lambda}(N) (\ap)^{\lambda-1},  \label{eq:B}
\end{equation}
and $g_i(N)$, $i=1$, 2, \ldots, $\lambda$, are some complex functions of $N$.
Equation~(\ref{eq:D-1}) is satisfied provided 
\begin{equation}
  \prod_{i=1}^{\lambda} g_i(N) = \prod_{i=1}^{\lambda} f_i(N), \label{eq:cond-g} 
\end{equation}
while Eq.~(\ref{eq:D-2}) imposes the conditions
\begin{eqnarray}
  f_i(N) g_{i+1}(N) & = & q f_{i+1}(N) g_i(N), \qquad i=1, 2, \ldots, \lambda-1,
       \nonumber \\
  f_{\lambda}(N) g_1(N) & = & q f_1(N) g_{\lambda}(N).  
\end{eqnarray}
The general solution of the latter is given by $g_i(N) = q^{i-1} f_i(N) k(N)$, $i=1$, 2,
\ldots, $\lambda$, in terms of some complex function $k(N)$, which, from
Eq.~(\ref{eq:cond-g}), is restricted by the condition $k^{\lambda}(N) = q^{-\lambda
(\lambda-1)/2}$. Up to some $N$-dependent $\lambda$th root of unity, which for
simplicity's sake we assume equal to 1, $k(N)$ is therefore obtained as $k(N) =
q^{-(\lambda-1)/2}$, so that the functions $g_i(N)$ are finally given by
\begin{equation}
  g_i(N) = q^{-(\lambda- 2i+1)/2} f_i(N), \qquad i=1, 2, \ldots, \lambda.  \label{eq:g}
\end{equation}
\par
%
%
It results from Eq.~(\ref{eq:h-bis}) and from the assumptions (\ref{eq:F}) and
(\ref{eq:phi(n)}) that the spectrum of $H$ is nonnegative. A complete set of
eigenvectors is given in terms of the Fock space basis states (\ref{eq:basis}) by
\begin{equation}
  |\phi_0, i\rangle = |i - d_{j-1} - 1\rangle e_{j,1}, \qquad i=1, 2, \ldots, \case{1}{2}
  \lambda (\lambda-1), \label{eq:phi-0}
\end{equation}
for $E_0 = 0$ and
\begin{equation}
  |\phi_n, i\rangle = |n + \lambda - 1 - i\rangle e_{i,1}, \qquad i=1, 2, \ldots, \lambda, 
  \label{eq:phi-n}
\end{equation}
for $E_n = \varphi(n+\lambda-2) \prod_{j=1}^{\lambda-1} F(n+\lambda-1-j) > 0$,
$n=1$, 2,~\ldots. In (\ref{eq:phi-0}), $j = j(i) \in \{1, 2, \ldots, \lambda-1\}$ is
determined by the condition $d_{j-1}+1 \le i \le d_j$, where $d_j \equiv j
(2\lambda-j-1)/2$.\par
%
%
All the excited states are therefore $\lambda$-fold degenerate and the fractional
supercharge $Q$ acts cyclically on them: $|\phi_n,1\rangle \to |\phi_n,2\rangle \to
\cdots \to |\phi_n,\lambda\rangle \to |\phi_n,1\rangle$. For the $\frac{1}{2}\lambda
(\lambda-1)$-fold degenerate ground state, the action of $Q$ is more complicated since
$Q |\phi_0, d_{j-1}+1\rangle = 0$, while for $d_{j-1}+2 \le i \le d_j$, $Q
|\phi_0,i\rangle \propto f_j(i+j-d_{j-1}-2) |\phi_0, i+\lambda-j-1\rangle$ may be
vanishing or not according to the value assumed by $f_j(i+j-d_{j-1}-2)$. Since
$i+j-d_{j-1}-2 \le \lambda-2$, condition (\ref{eq:phi(n)}) does not indeed ensure the
nonvanishing of the latter.\par
%
%
The $\lambda \times \lambda$-matrix realization (\ref{eq:Q-H}), (\ref{eq:D}) of $H$,
$Q$, and $D$ can be diagonalized through a unitary transformation $U = 
\sum_{i,j=1}^{\lambda} P_{i-j} e_{i,j}$, expressed in terms of the
projection operators $P_{\mu}$ defined in (\ref{eq:P}). The results read
\begin{eqnarray}
  H' & \equiv & U H U^{\dagger} = \diag(H_0, H_1, \ldots, H_{\lambda-1}),  \nonumber
         \\ 
  Q' & \equiv & U Q U^{\dagger} = \diag(Q_0, Q_1, \ldots, Q_{\lambda-1}), 
         \label{eq:diag} \\
  D' & \equiv & U D U^{\dagger} = \diag(D_0, D_1, \ldots, D_{\lambda-1}), \nonumber
\end{eqnarray}
where
\begin{equation}
  H_{\mu} = \sum_{i=1}^{\lambda} h_i(N) P_{\mu-i+1},  \qquad
  Q_{\mu} = \sum_{i=1}^{\lambda} A_i P_{\mu-i+1}, \qquad 
  D_{\mu} = \sum_{i=1}^{\lambda} B_i P_{\mu-i+1},  \label{eq:boson}
\end{equation}
for $\mu=0$, 1,~\ldots, $\lambda-1$, and $h_i(N)$, $A_i$, and $B_i$ are respectively
given by Eqs.~(\ref{eq:h-bis}), (\ref{eq:A}), and (\ref{eq:B}), together with
Eqs.~(\ref{eq:phi}) and (\ref{eq:g}). Each of the $\lambda$ sets of operators
$\{H_{\mu}, Q_{\mu}, D_{\mu}\}$ satisfies the FSSQM relations (\ref{eq:def-1}),
(\ref{eq:def-2}), (\ref{eq:D-1}), and (\ref{eq:D-2}), and is written in terms of a single
bosonic degree of freedom through the operators $N$, $\ap$, $a$ of \AG. We have
therefore proved that FSSQM of order $\lambda$ is fully reducible and we have obtained a
minimal bosonization thereof.\par
%
%
The eigenvalues $E^{(\mu)}_n$ of the bosonized fractional supersymmetric Hamiltonian
$H_{\mu}$, defined in (\ref{eq:boson}), can be written as
\begin{equation}
  E^{(\mu)}_{\lambda k + \nu} = \left\{\begin{array}{l}
      \varphi(\lambda k + \mu) \prod_{i=1}^{\lambda-1} F(\lambda k + \mu-i+1) \qquad
           {\rm if\ } \nu = 0, 1, \ldots, \mu, \\[0.2cm]
      \varphi[\lambda (k+1) + \mu] \prod_{i=1}^{\lambda-1} F[\lambda (k+1) + \mu-i+1]
           \\[0.2cm]
      \qquad{\rm if\ } \nu = \mu+1, \mu+2, \ldots, \lambda-1,
\end{array}\right.
\end{equation}
where $k=0$, 1,~\ldots. From Eqs.~(\ref{eq:F}) and (\ref{eq:phi(n)}), it follows that
\begin{eqnarray}
  E^{(\mu)}_0 & = & E^{(\mu)}_1 = \cdots =  E^{(\mu)}_{\mu} = 0, \nonumber \\ 
  E^{(\mu)}_{\lambda k+\mu+1} & = & E^{(\mu)}_{\lambda k+\mu+2} = \cdots =
       E^{(\mu)}_{\lambda (k+1)+\mu} > 0, \qquad k=0, 1, 2, \ldots,   
\end{eqnarray}
if $\mu=0$, 1, \ldots, or $\lambda-2$, and that 
\begin{equation}
  E^{(\lambda-1)}_{\lambda k} = E^{(\lambda-1)}_{\lambda k+1} = \cdots =
  E^{(\lambda-1)}_{\lambda (k+1)-1} > 0, \qquad k=0, 1, 2, \ldots,
\end{equation}
if $\mu = \lambda-1$.\par
%
%
In the former case, the corresponding eigenvectors may be written as
\begin{equation}
  \left|\phi^{(\mu)}_0, i\right\rangle = |\mu+1-i\rangle, \qquad i=1, 2, \ldots, \mu+1, 
  \label{eq:phi-mu-0}
\end{equation}
for $E = E^{(\mu)}_0 = 0$ and
\begin{equation}
  \left|\phi^{(\mu)}_k, i\right\rangle = |\lambda k +\mu+1-i\rangle, \qquad i=1, 2, \ldots,
  \lambda, 
\end{equation}
for $E = E^{(\mu)}_{\lambda k +\mu} > 0$, $k=1$, 2,~\ldots. Furthermore, $Q_{\mu} 
\left|\phi^{(\mu)}_0, i\right\rangle \propto f_i(\mu) \left|\phi^{(\mu)}_0,
i+1\right\rangle$, $i=1$, 2, \ldots,~$\mu$, and $Q_{\mu} 
\left|\phi^{(\mu)}_0, \mu+1\right\rangle = 0$, while $Q_{\mu}$ acts cyclically on
$\left|\phi^{(\mu)}_k, 1\right\rangle$, $\left|\phi^{(\mu)}_k, 2\right\rangle$, \ldots,
$\left|\phi^{(\mu)}_k, \lambda\right\rangle$ for $k=1$, 2,~\ldots. On the contrary, in
the latter case, one has
\begin{equation}
  \left|\phi^{(\lambda-1)}_k, i\right\rangle = |\lambda(k+1)-i\rangle, \qquad i=1, 2,
  \ldots, \lambda,  \label{eq:phi-k} 
\end{equation}
for $E = E^{(\lambda-1)}_{\lambda k} > 0$, $k=0$, 1, 2,~\ldots, and
$Q_{\lambda-1}$ has a cyclic action on all the sets of states
$\left|\phi^{(\lambda-1)}_k, 1\right\rangle$, $\left|\phi^{(\lambda-1)}_k,
2\right\rangle$, \ldots, $\left|\phi^{(\lambda-1)}_k, \lambda\right\rangle$.\par
%
%
We therefore conclude that in the present realization, for $\mu=0$ FSSQM is unbroken
with a nondegenerate ground state at a vanishing energy. For $\mu=1$, 2,~\ldots, or
$\lambda-2$, the ground state is still at a vanishing energy but is $(\mu+1)$-fold
degenerate and FSSQM is unbroken or broken according to whether $f_1(\mu) =
f_2(\mu) = \cdots = f_{\mu}(\mu) = 0$ or at least one of the $f_i(\mu)$, $i=1$, 2,
\ldots, $\mu$, is different from zero. Finally, for $\mu = \lambda-1$, the $\lambda$-fold
degenerate ground state lies at a positive energy and FSSQM is broken. In all the cases,
the excited states are $\lambda$-fold degenerate. It is worth noting that for the
standard realization of FSSQM in terms of ordinary bosonic operators and $q$-deformed
ones~\cite{durand}, only the counterparts of the $\mu=0$ case and of the $\mu =
\lambda-2$ one with broken FSSQM are obtained. The present realization therefore leads
to a much richer picture.\par
%
%
Before concluding this section, it is interesting to consider the $\lambda=2$ limit,
wherein FSSQM reduces to ordinary SSQM. In such a case, $q = \exp(\pi i) = -1$, so
that the $q$-commutator of Eq.~(\ref{eq:D-2}) becomes an anticommutator, the
functions $g_i(N)$ are given by $g_1(N) = - {\rm i} f_1(N)$, $g_2(N) = {\rm i}
f_2(N)$, and $Q$, $D$ are the usual supercharge and covariant derivative, respectively.
In the special case where $f_1(N) = f_2(N) = f(N)$ and $f(N)$ is a real function of $N$
such that $f(n) \in \R^+$ for $n \in \N^+$, $Q$ and $D$ are two Hermitian conserved
supercharges. From them, one can construct non-Hermitian ones,
\begin{eqnarray}
  {\cal Q} & \equiv & \frac{1}{\sqrt{2}} (Q + {\rm i}D) = \sqrt{2} \left(\begin{array}{cc}
       0 & 0 \\[0.1cm]
       f(N+1) a & 0
       \end{array}\right), \nonumber \\
  {\cal Q}^{\dagger} & \equiv & \frac{1}{\sqrt{2}} (Q - {\rm i}D) = \sqrt{2}
\left(\begin{array}{cc}
       0 & f(N) \ap \\[0.1cm]
       0 & 0
       \end{array}\right),
\end{eqnarray}
satisfying the usual SSQM defining relations ${\cal Q}^2 = ({\cal Q}^{\dagger})^2 = 0$,
$\{{\cal Q}, {\cal Q}^{\dagger}\} = H$, with $H = \diag(h_1(N), h_2(N))$ and $h_1(N)
= f^2(N) F(N)$, $h_2(N) = h_1(N+1)$. Such a realization of SSQM coincides with that
considered in Ref.~\cite{cq02} for PSSQM of order $p=1$.\par
%
%
\section{\boldmath $\mbox{\seventeenof Z}_{\lambda}$-Graded Uniform Topological
Symmetries of Type (1, 1, \ldots, 1) and Topological Invariants}

Before applying the concept of topological symmetries to the new realization of FSSQM
obtained in the previous section, let us briefly review the former.\par
%
%
According to Ref.~\cite{mosta}, a quantum system is said to possess a
\Z$_{\lambda}$-graded topological symmetry (TS) of type $(m_1, m_2, \ldots,
m_{\lambda})$ if and only if the following conditions are satisfied.
\begin{enumerate}

\item The quantum system is \Z$_{\lambda}$-graded. This means that the Hilbert space
$\cal H$ of the quantum system is the direct sum of $\lambda$ of its (nontrivial)
subspaces ${\cal H}_i$, $i=1$, 2, \ldots, $\lambda$, whose vectors are said to have a
definite grading $c_i$. In addition, the Hamiltonian $H$ of the system has a complete
set of eigenvectors with definite grading.

\item The energy spectrum is nonnegative.

\item For every positive energy eigenvalue $E$, there is a positive integer $d_E$ such
that $E$ is $d_E m$-fold degenerate and the corresponding eigenspaces are spanned by
$d_E m_1$ vectors of grade $c_1$, $d_E m_2$ vectors of grade $c_2$, \ldots, and
$d_E m_{\lambda}$ vectors of grade $c_{\lambda}$ (hence $m =
\sum_{i=1}^{\lambda} m_i$).
  
\end{enumerate}

\noindent 
One speaks of uniform topological symmetries (UTS) whenever $d_E = 1$ for all positive
energy eigenvalues $E$.\par
%
%
{}For a system with a \Z$_{\lambda}$-graded TS of type $(m_1, m_2, \ldots,
m_{\lambda})$, one can introduce a set of integer-valued topological invariants
$\Delta_{ij} \equiv m_i n^{(0)}_j - m_j n^{(0)}_i$, where $i$, $j \in \{1, 2, \ldots,
\lambda\}$ and $n^{(0)}_k$ denotes the number of zero-energy states of grade
$c_k$.\par
%
%
{}From the definition of TS, it is possible to obtain the underlying operator algebras
supporting such symmetries~\cite{mosta}. In particular, \Z$_2$-graded TS of type (1, 1)
has been shown to yield the SSQM algebra with $\Delta_{12}$ reducing to the Witten
index.\par
%
%
Here we shall restrict ourselves to a special case of \Z$_{\lambda}$-graded UTS of type
(1, 1, \ldots, 1), whose algebra coincides with that of FSSQM of order $\lambda$. For a
quantum system with Hamiltonian $H$ to have a \Z$_{\lambda}$-graded TS of type (1,
1, \ldots, 1), it is indeed sufficient that the following conditions be fulfilled.

\begin{enumerate}

\item There exist a grading operator $\tau$ and a TS generator $Q$ satisfying the
relations
\begin{equation}
  \tau^{\lambda} = 1, \qquad \tau^{\dagger} = \tau^{-1}, \qquad [H, \tau] = 0, \qquad
  [\tau, Q]_q = 0,  \label{eq:cond-tau}
\end{equation}
with $q = \exp(2\pi {\rm i}/\lambda)$, as well as Eqs.~(\ref{eq:def-1}) and
(\ref{eq:def-2}).

\item The spectrum of $H$ is nonnegative. 

\end{enumerate}

\noindent
The presence of this particular TS in turn implies the existence of $l = [\lambda/2]$
Hermitian operators $M_i$, $i=1$, 2, \ldots, $l$, commuting with $\tau$ and $Q$ and
fulfilling the equations
\begin{eqnarray}
  (Q_1^2 - M_1) (Q_1^2 - M_2) \ldots (Q_1^2 - M_l) & = & 2^{-l+1} H, \nonumber \\
  (Q_2^2 - M_1) (Q_2^2 - M_2) \ldots (Q_2^2 - M_l) & = & (-1)^l 2^{-l+1} H, \qquad
       {\rm if\ }\lambda = 2l,  \label{eq:M}
\end{eqnarray}
or
\begin{eqnarray}
  (Q_1^2 - M_1) (Q_1^2 - M_2) \ldots (Q_1^2 - M_l) Q_1 & = & 2^{-l+1/2} H,
       \nonumber \\
  (Q_2^2 - M_1) (Q_2^2 - M_2) \ldots (Q_2^2 - M_l) Q_2 & = & 0, \qquad
       {\rm if\ }\lambda = 2l+1,  \label{eq:M-bis}
\end{eqnarray}
where
\begin{equation}
  Q_1 = \frac{1}{\sqrt{2}}(Q + Q^{\dagger}), \qquad Q_2 = \frac{1}{{\rm i}\sqrt{2}}(Q
  - Q^{\dagger}). 
\end{equation}
\par
%
%
{}For the $\lambda \times \lambda$-matrix realization of FSSQM of order $\lambda$ in
terms of GDOA generators considered in Sec.~3, the Hilbert space $\cal H$ is the direct
sum of $\lambda$ copies of the GDOA Fock space $\cal F$: ${\cal H}_i = {\cal F}$,
$i=1$, 2, \ldots, $\lambda$. With the grading operator $\tau$ realized by the $\lambda
\times \lambda$ matrix
\begin{equation}
  \tau = \sum_{i=1}^{\lambda} q^i e_{i,i},  \label{eq:tau}
\end{equation}
a grade $c_i = q^i$ is assigned to the $i$th Fock space ${\cal H}_i$. It is straightforward
to check that $\tau$, as defined in (\ref{eq:tau}), satisfies Eq.~(\ref{eq:cond-tau}) with
$H$ and $Q$ as expressed in (\ref{eq:Q-H}) and that $\tau |\phi_0, i\rangle = q^j |\phi_0,
i\rangle$, $\tau |\phi_n, i\rangle = q^i |\phi_n, i\rangle$ for the energy eigenstates
(\ref{eq:phi-0}) and (\ref{eq:phi-n}), respectively. Since it has been shown in Sec.~3
that the spectrum of $H$ is nonnegative and that all the positive-energy eigenvalues
are $\lambda$-fold degenerate, it follows that all the conditions for having a
\Z$_{\lambda}$-graded UTS of type (1, 1, \ldots, 1) are fulfilled. The topological
invariants for the present system are
\begin{equation}
  \Delta_{ij} = - \Delta_{ji} = i-j, \qquad 1 \le i < j \le \lambda.
\end{equation}
\par
%
%
On using Eqs.~(\ref{eq:Q-H}) -- (\ref{eq:phi}) and (\ref{eq:h-bis}), we can also obtain an
explicit form for the operators $M_i$ of Eq.~(\ref{eq:M}) or (\ref{eq:M-bis}),
\begin{equation}
  M_i = \frac{1}{2} \sum_{j=1}^{\lambda} m_{ij}(N) e_{j,j}, \qquad m_{ij}(N) = 
  m_{i1}(N+j-1),  \label{eq:M-i}
\end{equation}
where $m_{i1}(N)$, $i=1$, 2, \ldots, $l$, are real solutions of a $l$th-degree algebraic
equation. For $\lambda=3$, 4, and 5, the latter are given by
\begin{eqnarray}
  m_{11}(N) & = & \alpha_1(N) + \alpha_2(N) + \alpha_3(N), \\
  m_{i1}(N) & = & \case{1}{2} [\alpha_1(N) + \alpha_2(N) + \alpha_3(N) + \alpha_4(N)
        + (-1)^i \delta(N)], \quad i=1, 2, \nonumber \\  
  \delta(N) & \equiv & \{[\alpha_1(N) + \alpha_2(N) + \alpha_3(N) + \alpha_4(N)]^2 
        \nonumber \\
  && \mbox{} - 4[\alpha_1(N) \alpha_3(N) + \alpha_2(N) \alpha_4(N)]\}^{1/2},
        \label{eq:m-4} \\
  m_{i1}(N) & = & \case{1}{2} [\alpha_1(N) + \alpha_2(N) + \alpha_3(N) + \alpha_4(N)
        + \alpha_5(N) + (-1)^i \delta(N)], \quad i=1, 2, \nonumber \\  
  \delta(N) & \equiv & \{[\alpha_1(N) + \alpha_2(N) + \alpha_3(N) + \alpha_4(N)
        + \alpha_5(N)]^2 - 4[\alpha_1(N) \alpha_3(N) 
        \nonumber \\
  && \mbox{}  + \alpha_2(N) \alpha_4(N) + \alpha_3(N) \alpha_5(N) + \alpha_4(N)
        \alpha_1(N) + \alpha_5(N) \alpha_2(N)]\}^{1/2}.  \label{eq:m-5}
\end{eqnarray} 
Here $\alpha_i(N) \equiv |f_i(N)|^2 F(N+1-i)$ if $i=1$, 2, \ldots, $\lambda-1$, and
$\alpha_{\lambda}(N) \equiv |f_{\lambda}(N)|^2 \prod_{j=1}^{\lambda-1} F(N+1-j)$. It
can be checked that the operators within the square roots in (\ref{eq:m-4}) and
(\ref{eq:m-5}) are nonnegative in $\cal F$ as it should be.\par
%
%
Let us finally consider the fully reduced form of FSSQM given in Eqs.~(\ref{eq:diag}) and
(\ref{eq:boson}). The transformed operators corresponding to $\tau$ and $M_i$, defined
in (\ref{eq:tau}) and (\ref{eq:M-i}), are given by
\begin{eqnarray}
  \tau' & \equiv & U \tau U^{\dagger} = \diag(\tau_0, \tau_1, \ldots, \tau_{\lambda-1}),
        \nonumber \\
  M'_i & \equiv & U M_i U^{\dagger} = \diag(M_{i,0}, M_{i,1}, \ldots, M_{i,\lambda-1}),
  \label{eq:diag-bis}
\end{eqnarray}
where 
\begin{equation}
  \tau_{\mu} = q^{\mu+1} T^{-1}, \qquad M_{i,\mu} = \sum_{j=1}^{\lambda} m_{ij}(N)
  P_{\mu-j+1},  \label{eq:boson-bis}
\end{equation}
for $\mu=0$, 1, \ldots, $\lambda-1$, and $T$ is defined in Eq.~(\ref{eq:T}). From
Eqs.~(\ref{eq:tau}), (\ref{eq:diag-bis}), and (\ref{eq:boson-bis}), it can also be shown
that $\tau' = \tau T^{-1}$.\par
%
%
{}For each $\mu$ value, $\tau_{\mu}$, $Q_{\mu}$, and $H_{\mu}$ satisfy the defining
assumptions of a \Z$_{\lambda}$-graded UTS of type (1, 1, \ldots, 1) with $\cal H$ now
coinciding with $\cal F$. We therefore get $\lambda$ realizations of such a UTS in the
same space $\cal F$, differing from one another by the grade assigned to the
subspaces ${\cal F}_{\nu}$, $\nu=0$, 1, \ldots, $\lambda-1$, which according to
(\ref{eq:T}) and (\ref{eq:boson-bis}), is given by $c^{(\mu)}_{\nu} = q^{\mu-\nu+1}$.
Hence, the subspaces ${\cal H}_1$, ${\cal H}_2$, \ldots, ${\cal H}_{\lambda}$ of $\cal
H$ with grade $q$, $q^2$, \ldots, $q^{\mu+1}$, $q^{\mu+2}$, \ldots,
$q^{\lambda-1}$, 1 are to be identified with ${\cal F}_{\mu}$, ${\cal F}_{\mu-1}$,
\ldots, ${\cal F}_0$, ${\cal F}_{\lambda-1}$,  \ldots, ${\cal F}_{\mu+2}$, ${\cal
F}_{\mu+1}$, respectively. For all the energy eigenstates (\ref{eq:phi-mu-0}) --
(\ref{eq:phi-k}), we then obtain $\tau_{\mu}
\left|\phi^{(\mu)}_k, i\right\rangle = q^i \left|\phi^{(\mu)}_k, i\right\rangle$, $k=0$, 1,
2,~\ldots. As a consequence, the topological invariants are now
\begin{equation}
  \Delta^{(\mu)}_{ij} = - \Delta^{(\mu)}_{ji} = \left\{\begin{array}{ll}
        -1 & \qquad {\rm if\ } 1 \le i \le \mu+1 < j \le \lambda, \\[0.2cm]
        0 & \qquad{\rm if\ } 1 \le i < j \le \mu+1 {\ \rm or\ }\mu+2 \le i < j \le \lambda,
  \end{array}\right.
\end{equation}
for $\mu=0$, 1, \ldots, $\lambda-2$, and
\begin{equation}
  \Delta^{(\lambda-1)}_{ij} = - \Delta^{(\lambda-1)}_{ji} = 0 \qquad {\rm if\ } 1 \le i < j
  \le \lambda,
\end{equation}
for $\mu = \lambda-1$.\par
%
%
\section{Conclusion}

In this letter, we have extended to FSSQM of order $\lambda$ the approach to PSSQM
and OSSQM in terms of GDOAs that we had previously proposed and we have obtained
both a fully reducible realization and a minimal bosonization of the theory. Furthermore,
we have provided some explicit examples of \Z$_{\lambda}$-graded UTS of type (1, 1,
\ldots, 1) and we have evaluated the corresponding topological invariants.\par
%
%
As in the cases of Beckers-Debergh PSSQM and of OSSQM, it turns out that in the limit
$G(N) \to I$, the fractional supersymmetric Hamiltonian and supercharge contain powers
of $P^2$ and $P$, respectively. Such features, characteristic of higher-derivative
SSQM~\cite{andrianov} and $\cal N$-fold SSQM~\cite{aoyama}, hint at possible
connections with such theories.\par
%
%

\end{document}